\documentclass[conference]{IEEEtran}
\IEEEoverridecommandlockouts

\usepackage{lscape}
\usepackage{rotating}
\usepackage{epstopdf}
\usepackage{etoolbox}
\usepackage{nomencl}
\usepackage{amsthm}
\usepackage{supertabular}
\usepackage{amsfonts}
\usepackage[dvips]{epsfig}
\usepackage{latexsym}
\usepackage{amssymb}
\usepackage{amsmath}
\usepackage{verbatim}
\usepackage{setspace}
\usepackage[ruled,linesnumbered,boxed,commentsnumbered]{algorithm2e}
\usepackage{bm}
\usepackage[colorlinks=true, linkcolor=black, citecolor=blue, urlcolor=blue]{hyperref}       
\usepackage{url}            
\usepackage{lipsum}		
\usepackage{cite}
\usepackage{amsmath,amssymb,amsfonts}
\usepackage{algorithmic}
\usepackage{graphicx}
\usepackage{textcomp}
\usepackage{xcolor}
\usepackage{hyperref}
\usepackage[figure]{hypcap}
\hypersetup{hypertex=true,
            colorlinks=true,
            linkcolor=blue,
            anchorcolor=blue,
            citecolor=blue}
\def\BibTeX{{\rm B\kern-.05em{\sc i\kern-.025em b}\kern-.08em
    T\kern-.1667em\lower.7ex\hbox{E}\kern-.125emX}}
    
\begin{document}

\title{Network-Aware Electric Vehicle Coordination for Vehicle-to-Anything Value Stacking Considering Uncertainties\\
\thanks{*Corresponding author: Hao Wang.}
\thanks{This work has been supported in part by the Australian Research Council (ARC) Discovery Early Career Researcher Award (DECRA) under Grant DE230100046.}
}

\author{\IEEEauthorblockN{Canchen Jiang\textsuperscript{1}, Ariel Liebman\textsuperscript{1,2}, Hao Wang\textsuperscript{1,2*}}
\IEEEauthorblockA{\textsuperscript{1} Department of Data Science and AI, Faculty of IT, Monash University, Australia \\
\textsuperscript{2} Monash Energy Institute, Monash University, Australia\\
Email: \texttt{\{canchen.jiang,ariel.liebman,hao.wang2\}@monash.edu}}
}

\maketitle

\begin{abstract}
The increased adoption of electric vehicles (EVs) has led to the development of vehicle-to-anything (V2X) technologies, including vehicle-to-home (V2H), vehicle-to-grid (V2G), and energy trading of EVs in the local grid. The EV coordination can provide value to the grid and generate benefits for EVs. However, network constraints and uncertainties in renewable energy and demand pose significant challenges to EV coordination and restrict the realization of these benefits. This paper develops a rolling-horizon optimization problem for V2X value stacking to fully unlock the value of EV coordination, considering power network constraints (such as voltage limits) and uncertainties in the energy system. By coordinating EVs to perform V2H, V2G, and energy trading, our approach exploits the most valuable services in real-time. We also analyze the expected extra costs caused by the prediction errors to evaluate the impact of uncertainties on V2X value stacking. We validate our value-stacking model using real data from Australia's National Electricity Market (NEM), ISO New England (ISO-NE), and New York ISO (NY-ISO) in the US. The results show that V2X value stacking achieves significant benefits to EVs through energy cost reduction. The uncertainty in the load has a higher impact on the value-stacking performance than PV generation, indicating the importance of load prediction.
\end{abstract}

\section{Introduction}
Electric vehicles (EVs) have become more popular worldwide, due to the advancements in battery technology and a strong commitment to a zero-emission transport future \cite{bibak2021comprehensive}. With an increased uptake of EVs, there has been significant progress in developing technologies for EVs to provide services to the grid, homes, and buildings through Vehicle-to-Anything (V2X) technology \cite{kester2018promoting},\cite{bitencourt2017optimal}. Residential EV coordination has been proposed to unlock of the value of EVs, but performing V2X often involves charge and discharge, resulting in voltage fluctuations in the distribution network. Moreover, the V2X performance can be affected by uncertainties in the energy system, including renewable energy generation and household energy consumption. Therefore, it is essential to develop effective methods to maximize the value of EVs while considering these uncertainties and maintaining power network constraints. 

Recent studies have explored methods to optimally schedule EV charging and discharging to interact with smart homes, local energy markets, and the grid to maximize the value of EVs. For example, Nakano et al. in \cite{nakano2020aggregation} studied the aggregation of vehicle-to-home (V2H) systems to participate in a regulation market using in-vehicle batteries. Bijan et al. in \cite{bibak2021influences} studied vehicle-to-grid (V2G) and EV energy trading in the local grid by analyzing the impact of V2G on reliability, cost, and emissions considering stochastic renewable energy. Shurrab et al. in \cite{shurrab2021efficient} proposed a vehicle-to-vehicle (V2V) energy-sharing framework to maximize social welfare while meeting the charging demand. Al-Obaidi et al. in \cite{al2020electric} designed an EV scheduling method to support peer-to-peer (P2P) energy trading and ancillary services. Thompson et al. in \cite{thompson2020vehicle} developed a vehicle-to-anything (V2X) value stream framework, including V2H, vehicle-to-building (V2B), and V2G, to study the economic potential of V2X. However, the multiple-value stream of EVs may be restricted by local network constraints, and the results are not convincing if such constraints are not considered. 

Power network constraints have been considered in more recent studies on EV scheduling and coordination. For example, Mazumder et al. in \cite{mazumder2020ev} studied the optimal EV charging/discharging to provision V2G and support voltage management in a distribution network. Affolabi et al. in \cite{affolabi2021optimal} proposed a two-level market framework for energy trading among EVs in charging stations taking into account power network constraints. Nevertheless, uncertainties in the energy system, such as stochastic renewable energy generation and consumers' time-varying load, pose significant challenges to EV scheduling, causing infeasible scheduling and network constraint violation. Even with advanced predictions for renewable generation and load, prediction errors are inevitable and will affect the performance of EV scheduling and coordination.

Real-time EV scheduling and coordination can mitigate the impact of uncertainties, and rolling horizon optimization (RHO) is an effective method for real-time decision-making. RHO has been used to mitigate the impact of uncertainties in the operation of energy systems \cite{kopanos2014reactive}. Su et al. in \cite{Su2020} proposed an RHO method for EVs in an electricity exchange market to tackle the uncertainty of electricity prices. Nimalsiri et al. in \cite{nimalsiri2021coordinated} proposed an RHO method to handle the uncertainty of EV arrival and departure time for a coordinated EV charging and discharging problem while maintaining the feeder voltage within safety limits. As illustrated above, prediction errors are inevitable, and prediction errors often have a negative impact on EVs enjoying the benefits from V2X services. The aforementioned studies did not consider how each uncertainty affects the system operation and value stacking benefits. In contrast, our work aims to maximize the value of EVs within power network constraints and quantify the impact of prediction errors of renewable energy and demand, highlighting the importance of different uncertainties in the value-stacking problem.

In this paper, we are motivated to develop a V2X value-stacking optimization problem to maximize the economic benefits for residential households with EVs, while maintaining the local grid voltage within limits under prediction errors. We adopt the RHO method to solve EV value-stacking decisions in real-time and evaluate the impact of prediction errors of solar PV generation and consumers' load. We validate our value-stacking optimization using real-world data from Australia's National Electricity Market (NEM), ISO New England (ISO-NE), and New York ISO (NY-ISO) in the US. The main contributions of this paper are as follows.
\begin{itemize}
    \item We develop a V2X value-stacking problem for residential EVs to exploit multiple value streams, including V2H, V2G, and energy trading, considering distribution network constraints and accounting for prediction errors. We formulate an RHO problem to solve the optimal value-stacking decisions for EV coordination in real-time based on predicted PV generation and energy demand of households using Long Short-Term Memory (LSTM).
    \item We evaluate the impact of prediction errors of PV generation and demand on the EV value-stacking optimization by analyzing the relationship between the extra cost and prediction errors. Our analysis shows that the uncertainty in household energy demand has a greater impact than the PV generation uncertainty, highlighting the importance of energy demand forecasting.
    \item We validate our V2X value-stacking method by considering two retail tariffs, e.g., time-of-use (TOU) pricing and two-part tariff (TPT) pricing, for EVs, and using real-world data from three markets, including Australia's NEM, ISO-NE, and NY-ISO in the US. The results show that, amongst all modeled value streams, V2H contributes most to the cost reduction in NEM and ISO-NE. Energy trading contributes most to the cost reduction in NY-ISO. Value-stacking leads to higher cost reduction under TOU pricing than TPT pricing. Energy trading creates the most marginal value compared to V2G and V2H in all three markets.

\end{itemize}
The remainder of this paper is organized as follows. Section \ref{Model system} presents the residential prosumer model with EVs and the power network model. Section \ref{Optimal Problem Formulation} formulates the rolling horizon value-stacking optimization problem and LSTM-based prediction model for PV generation and residential energy demand. Section \ref{evaluation} presents three baseline optimization problems and performance evaluation metrics. Section \ref{Simulation and Discussion} evaluates our proposed value-stacking method using real-world data from NEM, ISO-NE, and NY-ISO, respectively. Section \ref{Conclusion and Future Work} concludes our work.

\section{System Model}\label{Model system}
\begin{figure}[!t]
\centering
\includegraphics[width=0.99\columnwidth]{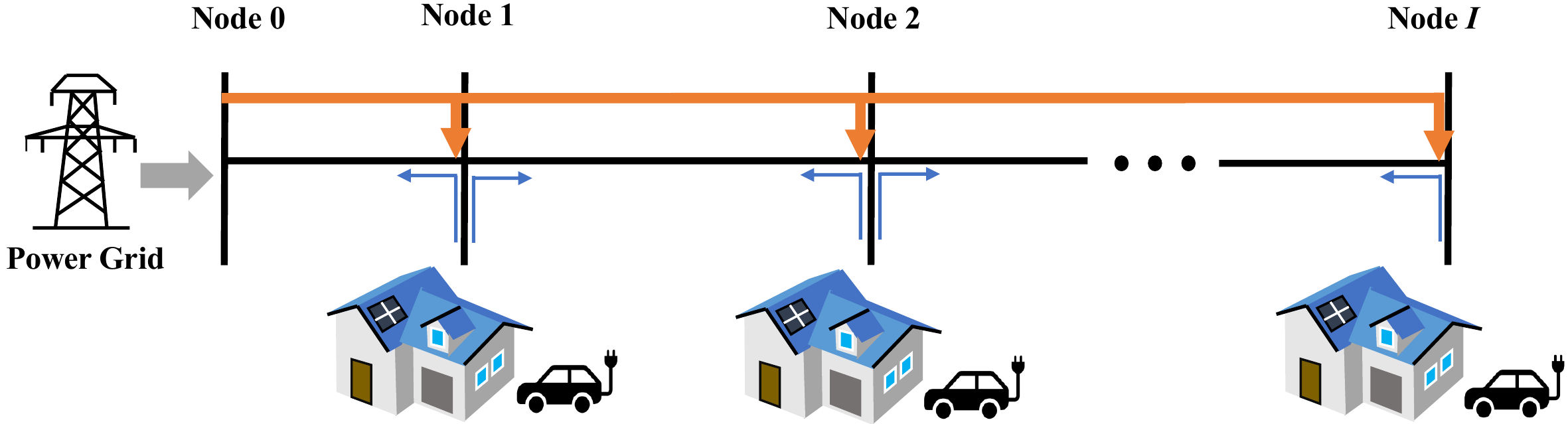}
\caption{Illustration of residential EV Coordination in a local network.}
\label{fig:system model}
\end{figure}

We consider an energy system with multiple households (as prosumers) having EVs connected through a local distribution network.
Fig. \ref{fig:system model} depicts an illustrative system for EV coordination, where households and EVs sit on a radial power network \cite{alskaif2019decentralized}. The orange bold line represents the power from the main grid, and the blue solid line represents the power flow of V2G and energy trading in the local network. We define $\mathcal{I}=\left\{1,\dots, I\right\}$ as the set of nodes in the distribution network. Additionally, we denote $\mathcal{U}=\left\{1,\dots, N\right\}$ as the set of prosumers, and $N$ is the total number of prosumers. Among prosumers, there is a group of prosumers connected to node $i$, denoted as $\mathcal{B}^{i} \subseteq \mathcal{U}$. It is noted that each prosumer must belong to one group and the prosumer groups $\mathcal{B}^{1}, \mathcal{B}^{2}, \dots, \mathcal{B}^{I}$ do not overlap with each other, i.e., $\mathcal{B}^{1} \cup \mathcal{B}^{2} \cup \dots \cup \mathcal{B}^{I} = \mathcal{U}$ and $\mathcal{B}^{1} \cap \mathcal{B}^{2} \cap \dots \cap \mathcal{B}^{I} = \emptyset$. We denote the operational horizon as $\mathcal{H}=\left\{1,\dots, H\right\}$ with evenly-spaced time slots, and consider $H=24$ hours for a day. 
 
\subsection{Prosumer Model (EV, Load, and Supply)}
Prosumers park their EVs at home during a time window denoted as $[t^{u}_m,t^{u}_n]\subseteq {\mathcal{H}}$ for prosumer $u$ and need to charge the EVs to the desired amount $B_{\text{EVdes}}^{u}$ for travel at $t_n$. Additionally, we consider EVs need to be charged to the requested energy $B_{\text{EVres}}^{u}$ in the last time slot to ensure more realistic scenario. The EV battery dynamics of prosumer $u$ at $t$ is
\begin{align}\label{eq1}
    &b_{\text{EVs}}^{u,t} = b_{\text{EVs}}^{u,t-1} + \mu p_{\text{EVc}}^{u,t} - \frac{1}{\eta}p_{\text{EVd}}^{u,t},~t\in {[t^{u}_m,t^{u}_n]},\\
    &b_{\text{EVs}}^{u,t} = B_{\text{EVdes}}^{u},~t= t^{u}_n,
\end{align}
where $\mu\in [0,1]$ and $\eta\in (0,1]$ denote the charging and discharging efficiency of prosumer $u$'s EV, respectively. The charge power and discharge power are denoted as $p_{\text{EVc}}^{u,t}$ and $p_{\text{EVd}}^{u,t}$ in $t$-th time slot, and $b_{\text{EVs}}$ is the stored energy in prosumer $u$'s EV. 

The EV energy $b_{\text{EVs}}$ is constrained as
\begin{equation}\label{eq2}
    \underline{B}_{\text{EVs}}^{u} \leq b_{\text{EVs}}^{u,t} \leq \overline{B}_{\text{EVs}}^{u},
\end{equation}
where $\overline{B}_{\text{EVs}}$ and $\underline{B}_{\text{EVs}}$ are the upper and lower bounds.

The constraints of EV charging and discharge are
\begin{align}
    0 \leq p_{\text{EVc}}^{u,t} \leq \overline{P}_{\text{EVc}}^{u,t} (1-x^{u,t}),~t\in {[t^{u}_m,t^{u}_n]}, \label{eq3}\\
    0 \leq p_{\text{EVd}}^{u,t} \leq \overline{P}_{\text{EVd}}^{u,t} x^{u,t},~t\in {[t^{u}_m,t^{u}_n]}, \label{eq4}
\end{align}
where $\overline{P}_{\text{EVc}}^{u,t}$ and $\overline{P}_{\text{EVd}}^{u,t}$ are the maximum limits for charging and discharging, respectively. Since EVs cannot charge and discharge at the same time slot, we introduce a binary variable $x^{u,t}\in \{ 0, 1\}$ to limit its operation. Note that if EVs are not parked, i.e., $t \notin {[t^{u}_m,t^{u}_n]}$, $p_{\text{EVc}}^{u,t} = 0$ and $p_{\text{EVd}}^{u,t} = 0$.

The EV charging/discharging incurs degradation of the battery, and the cost of EV battery degradation is 
\begin{equation}
   \mathbf{C}_{\text{battery}}^u = \alpha_b \sum \nolimits_{t\in \mathcal{H}} \left( (p_{\text{EVd}}^{u,t})^2+(p_{\text{EVc}}^{u,t})^2 \right), 
\end{equation}
where $\alpha_b$ is the amortized cost coefficient of the battery use \cite{wang2016incentivizing}.

EVs can be charged using local renewable (such as solar) $p_{\text{renew}}^{u,t}$, from local market $p_{\text{Buy}}^{u,t}$ and the grid $p_{\text{grid}}^{u,t}$, which satisfy
\begin{align}\label{eq5}
    0 \leq p_{\text{renew}}^{u,t} \leq R^{u,t}, \\
    0 \leq p_{\text{Buy}}^{u,t} \leq \overline{P}_{\text{Buy}}^{u,t},\\
    0 \leq p_{\text{grid}}^{u,t} \leq G^{u},
\end{align}
where $R^{u,t}$, $\overline{P}_{\text{Buy}}^{u,t}$, and $G^{u}$ are the upper-bounds of renewable energy, power prosumer $u$ purchase from the local market, and grid power for prosumer $u$ at $t$.

We evaluate two types of tariffs for prosumers as follows.
\begin{enumerate}
    \item Two-part tariff pricing (TPT): Prosumer $u$'s electricity cost is $$
    \mathbf{C}^{u}_{\text{grid}} = \pi_g\sum_{t\in \mathcal{H}} p_{\text{grid}}^{u,t} + \pi_{\text{peak}}\max_{t\in \mathcal{H}} p_{\text{grid}}^{u,t},$$ 
    where $\pi_\text{g}$ and $\pi_{\text{peak}}$ are fixed energy price and peak price respectively. TPT is used to incentivize peak shaving.
    \item Time-of-use pricing (TOU): Prosumer $u$'s electricity cost is $$
    \mathbf{C}^{u}_{\text{grid}} = \sum_{t\in \mathcal{H}} \pi_g^t p_{\text{grid}}^{u,t},$$ 
    where $\pi_g^t$ is the TOU prices in peak, off-peak, and shoulder hours.
\end{enumerate}

The demand of prosumer $u$ at time slot $t$ is met by various power supplies, shown as
\begin{equation}\label{eq7}
        p_{\text{\text{grid}}}^{u,t} + p_{\text{renew}}^{u,t} + p_{\text{Buy}}^{u,t} + p_{\text{V2H}}^{u,t} = P_{\text{Load}}^{u,t} + p_{\text{EVc}}^{u,t},
\end{equation}
where the left-hand side corresponds to the power supply from the grid $p_{\text{\text{grid}}}^{u,t}$, local renewable energy $p_{\text{renew}}^{u,t}$, the local energy market $p_{\text{Buy}}^{u,t}$, and V2H $p_{\text{V2H}}^{u,t}$. The right-hand side represents the total load demand including the inflexible load $P_{\text{Load}}^{u,t}$ and the EV charging demand $p_{\text{EVc}}^{u,t}$. Note that prosumer $u$ can also sell power to other prosumers and the grid, and we will introduce it in Section~\ref{valuestream}.

\subsection{Power Network Model}
For the power network, we consider a linearized distribution network model, which has been widely used to model distribution network constraints for active power, reactive power, and voltage \cite{wang2015decentralized,zhong2018topology}. The linearized power network model is as follows
\begin{align}
    &p^{i+1,t} = p^{i,t} - p_{\text{Load}}^{i,t}\label{eq16},\\
    &p_{\text{Load}}^{i,t} = P_{\text{Inflex}}^{i,t} + \sum_{u \in \mathcal{B}^{i}} (p_{\text{grid}}^{u,t} + p_{\text{Buy}}^{u,t} -(p_{\text{Sell}}^{u,t} + p_{\text{V2G}}^{u,t}))\label{eq17}, \\
    &P_{\text{DN}}^{\min} \leq  p_{\text{Active}}^{i,t} \leq P_{\text{DN}}^{\max}\label{eq18},\\
    & q^{i+1,t}  = q^{i,t} - Q_{\text{Load}}^{i,t}\label{eq19},\\
    & Q_{\text{DN}}^{\min} \leq  q_{\text{Reactive}}^{i,t} \leq Q_{\text{DN}}^{\max}\label{eq20},\\
    & v^{i+1,t}  =  v^{i,t} - \left( r_{i+1} p^{i+1,t} + x_{i+1} q^{i+1,t} \right) / V^{0,t}\label{eq21},\\
    &V^{i}_{\min} \leq  v^{i,t} \leq V^{i}_{\max}\label{eq22},
\end{align}
where $p^{i,t}$ is the active power flow from node $i-1$ to $i$ in time slot $t$, $p_{\text{Load}}^{i,t}$ is the amount of active power that consumed by both the active inflexible load $P_{\text{Inflex}}^{i,t}$ and the electricity usage of prosumers belonging to the set $\mathcal{B}^{i}$ in node $i$ at $t$. Note that $p_{\text{Load}}^{i,t}$ can have a negative value when prosumers sell power to the network from node $i$ at $t$. The maximum and minimum allowed active power are $P_{\text{DN}}^{\max}$ and $P_{\text{DN}}^{\min}$. The reactive power flow is denoted by $q^{i,t}$ from node $i-1$ to $i$ at $t$, and $Q_{\text{Load}}^{i,t}$ is reactive load at node $i$ in $t$. The voltage of node $i$ at $t$ is denoted as $v^{i,t}$, $V^{0,t}$ is the voltage of node $0$ at $t$, $r_i$ and $x_i$ is the resistance and reactance of branch connected to node $i-1$ and $i$ respectively, and $V^{i}_{\max}$ and $V^{i}_{\min}$ are the maximum and minimum value of voltage in each node $i$.

\section{Value-Stacking Optimization Problem Formulation}\label{Optimal Problem Formulation}
\begin{figure}[!b]
\centering
\includegraphics[width=0.99\columnwidth]{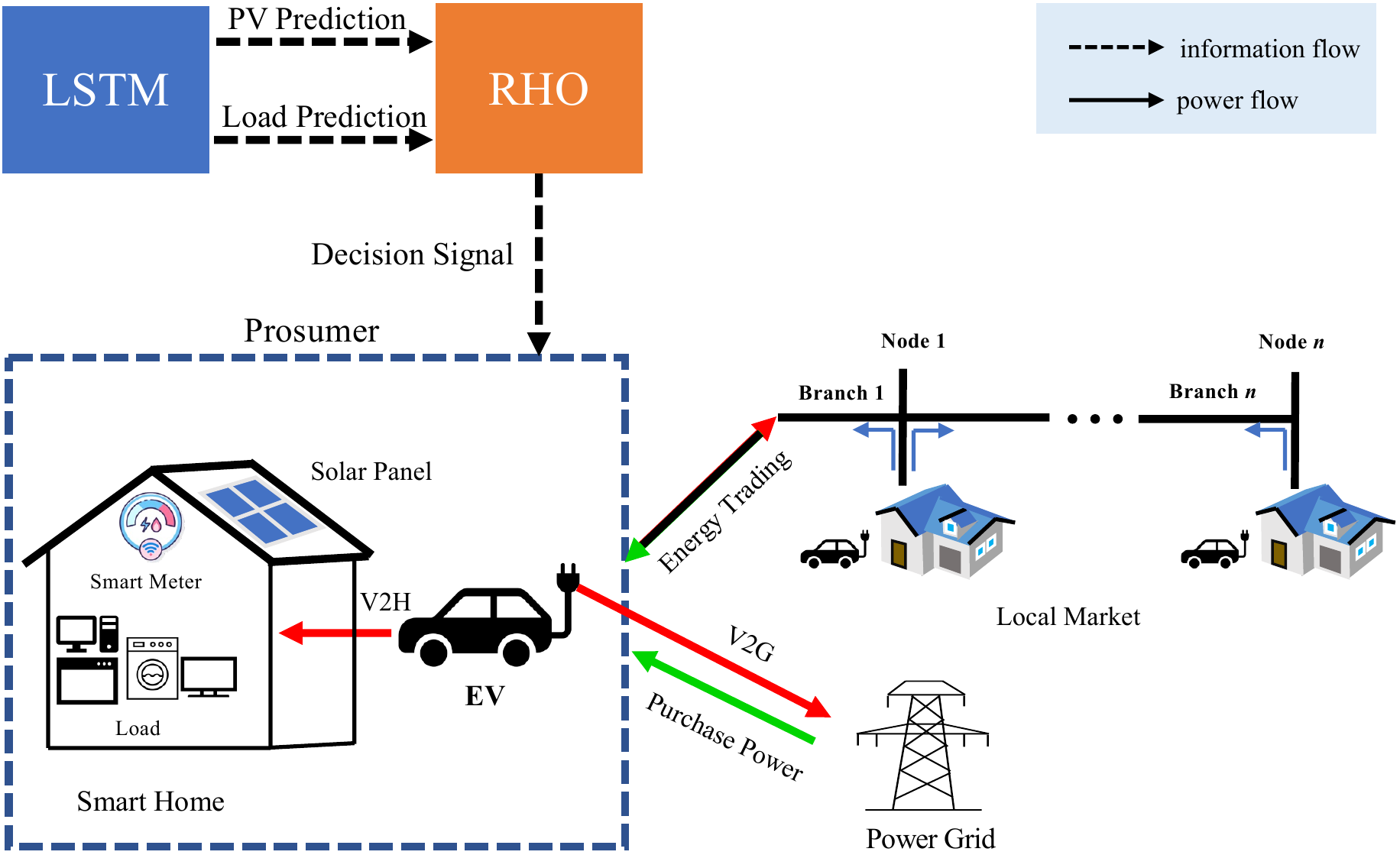}
\caption{Framework of the proposed value-stacking problem.}
\label{fig:valuestacking}
\end{figure}

In this section, we will introduce different value streams and formulate the value-stacking optimization problem. We illustrate the proposed value stacking problem in Fig.~\ref{fig:valuestacking}, where each prosumer has a solar PV system, an EV, load, and a smart meter with energy management functions. EVs can discharge the battery to perform V2H, V2G, and energy trading, and can be recharged from the main grid and PV. We develop a predict-and-optimize method. In each time slot $t$, we use an LSTM model to predict the household load and PV generation during the time slot $t$+1 to $H$ for prosumer $u$. Using the predicted data, we develop an RHO problem to solve the optimal value-stacking decision for prosumer $u$ over the operational horizon $[t+1, H]$. In each time slot $t$, Prosumer $u$ will implement the real-time decision in $t$ and repeat this process in a rolling manner.

\subsection{Value-Stacking Model}\label{valuestream}
We consider multiple value streams enabled by emerging technologies, including V2H, V2G, and energy trading in the local market. The prosumer optimizes its EV scheduling, especially discharging, to leverage the multiple value streams. The EV discharging satisfies the following constraint:
\begin{equation}\label{eq6}
    p_{\text{EVd}}^{u,t} =  p_{\text{V2H}}^{u,t} + p_{\text{V2G}}^{u,t} + p_{\text{Sell}}^{u,t},
\end{equation}
where $p_{\text{V2G}}^{u,t}$ and $p_{\text{Sell}}^{u,t}$ represent EV discharge to perform V2G, and energy trading in the local market, respectively. 

\subsubsection{Value Stream of V2H}
V2H is a technology enabled by the bidirectional charger for EVs to exchange energy with homes \cite{liu2013opportunities}. In V2H, EVs can store energy when the tariffs are low and supply power to homes to reduce the energy cost when the tariffs are high. The constraint of $p_{\text{V2H}}^{u,t}$ is 
\begin{equation}\label{eq9}
    0 \leq p_{\text{V2H}}^{u,t} \leq \overline{P}_{\text{V2H}}^{u,t},
\end{equation}
where $\overline{P}_{\text{V2H}}^{u,t}$ denotes the maximum allowed V2H power at $t$.

\subsubsection{Value Stream of V2G}
V2G is a technique to utilize EV batteries to sell energy back to the grid and provide ancillary services, such as spinning reverse \cite{machlev2020review}. 

In this section, we consider the revenue of V2G consisting of two parts. One part is that prosumer $u$ sells energy back to grid to make a profit. The revenue is $\pi_{\text{v2g}}^{t} p_{\text{V2G}}^{u,t}$, where $\pi_{\text{v2g}}^t$ is the dynamic price of electricity market in time slot $t$, $p_{\text{V2G}}^{u,t}$ is the V2G energy that prosumer $u$ sells to the grid. The constraint of $p_{\text{V2G}}^{u,t}$ is
\begin{equation}\label{eq10}
    0 \leq p_{\text{V2G}}^{u,t} \leq \overline{P}_{\text{V2G}}^{u,t},
\end{equation}
where $\overline{P}_{\text{V2G}}^{u,t}$ is the maximum V2G energy for $u$ at $t$.

The other part is EV ancillary service for the grid. We define $p_{\text{AS}}^{u,t}$ as the amount of reserve provided by prosumer $u$ using her EV battery in time slot $t$, and $\pi_{AS}^{t}$ is the unit price of reserve. Hence, the revenue of ancillary service (e.g., reserve) is $\pi_{\text{AS}}^{t} p_{\text{AS}}^{u,t}$. 

The total reserve of V2G for prosumer $u$ is 
$$
\mathbf{R}_{\text{V2G}}^{u,t} = \sum \nolimits_{t\in\mathcal{H}} \left( \pi_{\text{V2G}}^{t} p_{\text{V2G}}^{u,t} + \pi_{\text{AS}}^{t}p_{\text{AS}}^{u,t} \right).
$$ 

Since the reserve takes a part of the EV battery, and the bounds for stored energy in \eqref{eq2} are
\begin{align}
      &\overline{B}_{\text{EVs}}^{u} = B_{\text{EVs}}^{u,\max} - p_{\text{AS}}^{u,t}\label{eq11},\\
      &\underline{B}_{\text{EVs}}^{u} = B_{\text{EVs}}^{u,\min} + p_{\text{AS}}^{u,t}\label{eq12},
\end{align}
where $B_{\text{EVs}}^{u,\max}$ and $B_{\text{EVs}}^{u,\min}$ are maximum and minimum operational limits for prosumer $u$'s EV battery, respectively. In addition, the committed reserve $p_{\text{AS}}^{u,t}$ satisfies
\begin{equation}\label{eq13}
    0 \leq p_{\text{AS}}^{u,t} \leq B_{\text{EVs}}^{u,\max} /2.
\end{equation}

\subsubsection{Value Stream of Energy Trading in Local Market}
Prosumers can exchange power in the local market using their EVs. The dynamic prices are determined by the mid-market rate, which takes the mid-way value between the selling price and buying price \cite{long2017peer}.
We denote $\pi_{\text{Buy}}^{u,t}$ and $\pi_{\text{Sell}}^{u,t}$ as the buying and selling prices for each prosumer $u$ in the local market, respectively. Therefore, prosumer $u$'s profit and cost for energy trading in the local market are 
\begin{align*}
\mathbf{C}_{\text{ET}}^u = \sum_{t\in \mathcal{H}} \pi_{\text{Buy}}^{u,t} p_{\text{Buy}}^{u,t} - \sum_{t\in \mathcal{H}} \pi_{\text{Sell}}^{u,t} p_{\text{Sell}}^{u,t},
\end{align*}
where $\mathbf{C}_{\text{ET}}^u $ represent the cost of buying power minus revenue of selling power to the local market for each prosumer $u$. Since the each prosumer $u$ cannot sell power to local market and buy from it at the same time, we introduce binary variables $y^{u,t}\in \{ 0, 1\}$, and the constraints for energy exchange are 
\begin{align}
    &0 \leq p_{\text{Sell}}^{u,t} \leq \overline{P}_{\text{Sell}}^{u} y^{u,t}, \label{eq14}\\
    &0 \leq p_{\text{Buy}}^{u,t} \leq \overline{P}_{\text{Buy}}^{u}(1-y^{u,t}),\label{eq15}
\end{align}
where $\overline{P}_{\text{Sell}}^{u}$ and $\overline{P}_{\text{Buy}}^{u}$ denote the upper bounds for selling and buying energy in the local market for prosumer $u$. The total energy selling and buying in the local energy market should satisfy the balance constraint in each time slot $t$ as 
\begin{equation}\label{eq24}
    \sum_{u \in \mathcal{U}}p_{\text{Sell}}^{u,t} =  \sum_{u \in \mathcal{U}}p_{\text{Buy}}^{u,t}.
\end{equation}

\subsection{Load Demand and PV Generation Forecasting}\label{LSTM}
Residential load and solar PV generation forecasting will be needed in the value-stacking RHO optimization, presented in Section~\ref{valuestacking}. The forecasting would affect the scheduling for prosumers. Long Short-Term Memory (LSTM) is a type of recurrent neural network, which has been shown to be promising in time series forecasting \cite{kong2017short}. 
\begin{figure}[!t]
\centering
\includegraphics[width=0.46\columnwidth]{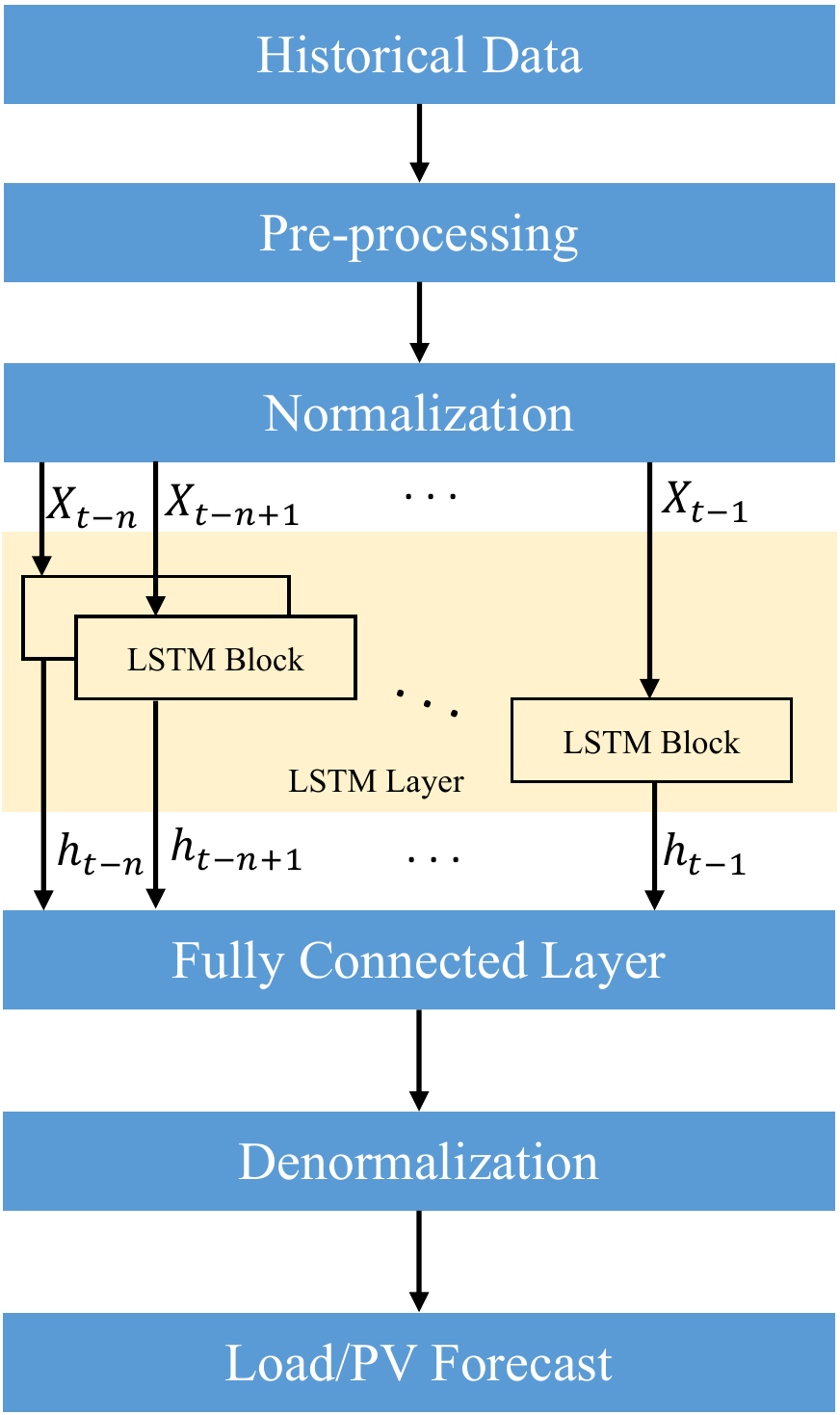}
\caption{Architecture of the LSTM model for predicting residential load and PV generation of households.}
\label{fig:LSTM}
\end{figure}

For residential load and PV generation time series forecasting, we utilize the LSTM network model to predict hourly load and PV generation, respectively. The architecture of the LSTM model is presented in Fig.~\ref{fig:LSTM}, where it starts with pre-processing for the input data, e.g., historical residential load and PV generation data. We denote $X = \{x_{1},x_{2},x_{3},\cdot\cdot\cdot ,x_{T}\}$ as the input sequence for the LSTM model, where $x_{t} \in \mathbf{R}^{n} $ is a n-dimensional vector of real values at the $t$-th time slot. To prevent the LSTM model training from diverging, we normalize the training sequence to have zero mean and unit variance. Additionally, for training a straightforward one-layer LSTM network, one needs to set the hyper-parameter of the hidden output dimension $n$. The hidden output $h_{t}$ at a specific time slot, in this work, is an n-dimensional vector. The sigmoid function is deployed in the fully connected layer to limit the output range from 0 to 1. After implementing denormalization, we can obtain the forecast results for load and PV generation.

\subsection{Rolling Horizon Value-Stacking Optimization Problem}\label{valuestacking}
\begin{figure}[!t]
\centering
\includegraphics[width=\columnwidth]{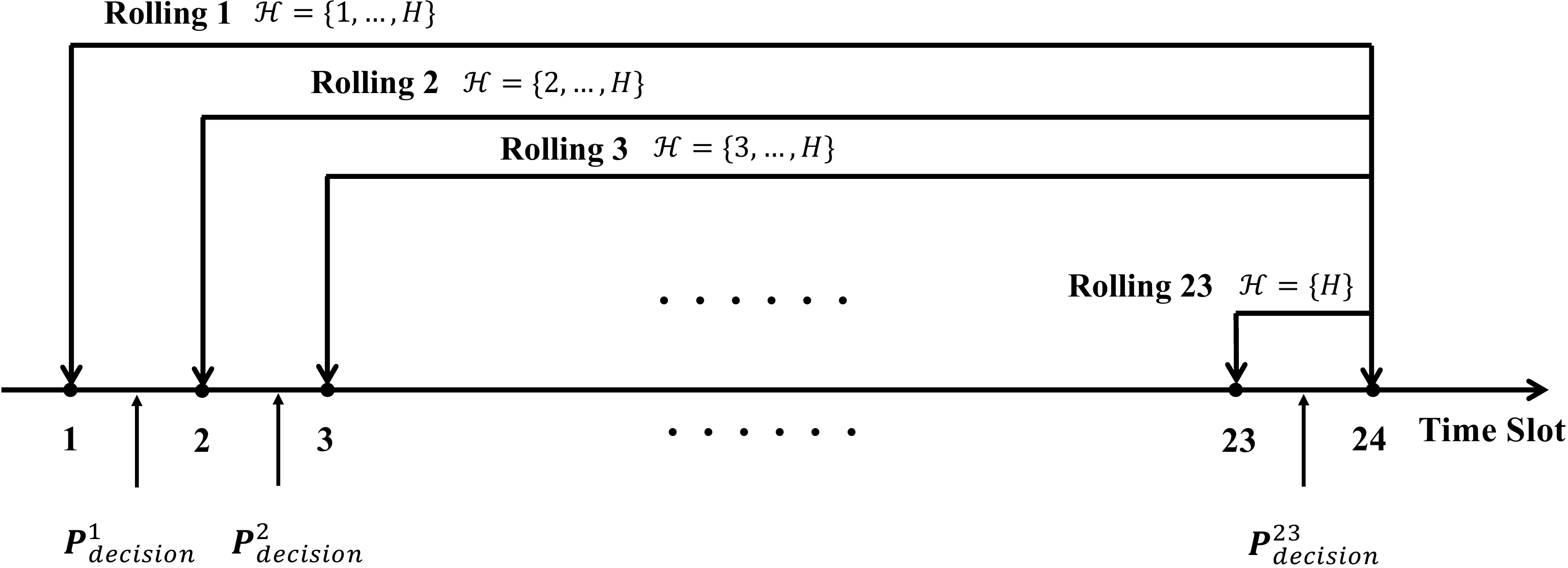}
\caption{The dynamic time window for the rolling-horizon EV value-stacking optimization.}
\label{fig:dynamic window}
\end{figure}

Given the system model in Sections~\ref{Model system}, value streams in Sections~\ref{valuestream}, and forecast of residential load and PV generation and detailed system model in Sections~\ref{LSTM}, we formulate the RHO problem for the real-time value-stacking problem. Various value streams, including V2H, V2G, and energy trading in the local grid, are co-optimized to schedule EVs. We let $\mathbf{C}_{S1}^{\text{total}}$ denote the total cost of all prosumers, and the objective is to minimize the total cost by scheduling variables, including $p_{\text{grid}}^{u,t}$, $p_{\text{renew}}^{u,t}$, $p_{\text{EVc}}^{u,t}$, $p_{\text{AS}}^{u,t}$, $b_{\text{EVs}}^{u,t}$, $p_{\text{V2H}}^{u,t}$, $p_{\text{V2G}}^{u,t}$, $p_{\text{Sell}}^{u,t}$, $p_{\text{Buy}}^{u,t}$, and $p_{\text{output}}^{u,t}$. The operational horizon $\mathcal{H}$ changes according to the dynamic time window presented in Fig. \ref{fig:dynamic window} to enable RHO. Hence, the optimization problem of value stacking is formulated as
\begin{equation}
\begin{aligned} \label{S1}
\min ~& \mathbf{C}_{S1}^{\text{total}} = \sum_{u \in \mathcal{U}}(\mathbf{C}^{u}_{\text{grid}}+\mathbf{C}_{\text{battery}}^u+\mathbf{C}_{\text{Buy}}^u-\mathbf{R}_{\text{Sell}}^u-\mathbf{R}_{\text{V2G}}^u)\\
\text{s.t.} ~& \eqref{eq1}-\eqref{eq4}~\text{and}~ \eqref{eq5}-\eqref{eq24}.
\end{aligned}
\end{equation}
  
The rolling-horizon value-stacking optimization problem is a mixed-integer Quadratic programming (MIQP) problem over the dynamic time window, as shown in Fig. \ref{fig:dynamic window}. The decisions for the current time slot will be executed. The process is presented in Algorithm~\ref{Process}, where we define $D = \{1,\dots, d\}$ as a set of operation days. At each time slot $t$, the EV value-stacking optimization solves the optimal decisions and generates a set of schedules for all prosumers based on the predicted load and PV generation from $t+1$ up to the end of the horizon $T$. In this process, only the decision variables in the current time slot are realized and used to calculate the actual cost. With rolling-horizon moving onward, the size of the dynamic window decreases by a one-time slot until the last time slot in the operational horizon. 
We take the optimal cost from the value-stacking problem with perfect predictions as the baseline to evaluate the impact of prediction errors. 

\begin{algorithm}[!t]
\DontPrintSemicolon
\caption{Rolling-horizon Optimization Solution}\label{Process}
\SetKwInOut{Input}{input}
\SetKwInOut{Output}{output}
\renewcommand{\Input}{\textbf{Initialization:}}
\Input{$\text{ Operation Day}~$d$ \leftarrow 1$;\;
Dynamic window $H$ $\leftarrow$ 24;\;
Time slot $t$ $\leftarrow$ 1;\;}
\For{$d = 1:D$}{\For{$H= t:24$}{\text{$\bullet$ Input realized load and PV generation data in $t$} \text{and predicted data from the LSTM model} \text{in time slots $[t+1,24]$ into Eq.~\eqref{S1};}\;

\text{$\bullet$ Save decision variable $\mathbf{P}_{\text{decision}}^{t}$ in time slot $t$;}\; 

\text{$\bullet$ Calculate the cost of all prosumers in current} \text{time slot $t$ based on $\mathbf{P}_{\text{decision}}^{t}$;}\;

\text{$\bullet$ Compute the sum of costs over all time slots} \text{according to }$\mathbf{C}_{\text{Total}}^{d} = \sum _{t \in \mathcal{H}}\mathbf{C}_{\text{Cost}}^{t,H};$\; 
$\bullet~ t \leftarrow t + 1$.}

\text{Save the $d$-th day data in $d$-th column}\;\text{in vector~}$\mathbf{C}_{\text{Total}} = [\mathbf{C}_{\text{Total}}^{1},\mathbf{C}_{\text{Total}}^{2},\cdots,\mathbf{C}_{\text{Total}}^{d}];$} 

\Output{Minimized cost for all prosumers based on the rolling-horizon optimization.}
\end{algorithm}

\section{Model Evaluation}\label{evaluation}
In this section, we present three baseline optimization problems for comparison with our value-stacking model. We also introduce the performance metrics for investigating the impact of prediction errors by measuring the extra cost based on the rolling-horizon optimization.

\subsection{Baseline Optimization Problems}\label{baseline}
To evaluate the contribution of each value stream, we study a set of baseline optimization problems. Specifically, we consider V2G, V2H, and EV energy trading alone, respectively, as three baseline problems.
\begin{itemize}
    \item For V2G alone, prosumers optimize V2G by scheduling variables, including $p_{\text{grid}}^{u,t}$, $p_{\text{renew}}^{u,t}$, $p_{\text{EVc}}^{u,t}$, $p_{\text{AS}}^{u,t}$, $b_{\text{EVs}}^{u,t}$, and $p_{\text{V2G}}^{u,t}$, where $p_{\text{EVd}}^{u,t} = p_{\text{V2G}}^{u,t}$.
    \item For V2H alone, prosumers minimize their costs by scheduling the variables, including $p_{\text{grid}}^{u,t}$, $p_{\text{renew}}^{u,t}$, $p_{\text{EVc}}^{u,t}$, $b_{\text{EVs}}^{u,t}$, and $p_{\text{V2H}}^{u,t}$, where $p_{\text{EVd}}^{u,t} = p_{\text{V2H}}^{u,t}$. 
    \item For EV energy trading alone, prosumers with EVs trade energy with others by scheduling variables, including $p_{\text{grid}}^{u,t}$, $p_{\text{renew}}^{u,t}$, $p_{\text{EVc}}^{u,t}$, $b_{\text{EVs}}^{u,t}$, $p_{\text{Sell}}^{u,t}$, and $p_{\text{Buy}}^{u,t}$, where $p_{\text{EVd}}^{u,t} = p_{\text{Sell}}^{u,t}$.
\end{itemize}

Moreover, to evaluate the marginal contribution of each value stream in our value-stacking problem presented in Section~\ref{valuestacking}, we consider three additional baseline problems. Each baseline problem excludes one value stream from the value-stacking problem with all value streams in Section~\ref{valuestacking}, showing the marginal contribution of each excluded value stream.

\subsection{Performance Metrics}
Compared to the optimal baseline cost, the relative extra cost (REC) caused by prediction errors is computed by 
\begin{equation}\label{eq26}
    \text{REC} = \frac{\sum_{t\in [t^{u}_m,t^{u}_n]} \sum_{u \in \mathcal{U}} |\widetilde{\mathbf{C}}_{\text{cost}}^{u,t} - \mathbf{C}_{\text{cost}}^{u,t}|}{\sum_{t\in [t^{u}_m,t^{u}_n]} \sum_{u \in \mathcal{U}} (\mathbf{C}_{\text{cost}}^{u,t})},
\end{equation}
where $\mathbf{C}_{\text{cost}}^{u,t}$ and $\widetilde{\mathbf{C}}_{\text{cost}}^{u,t}$ represent the cost of prosumer $u$ in time slot $t$ based on real-world data and prediction data, respectively.

For the forecasting model, e.g., LSTM in our work as an example, we calculate the relative errors (RE) in the predicted load and PV generation according to
\begin{align}
  & \text{RE}_{\text{load}} = \frac{\sqrt{\sum_{t\in [t^{u}_m,t^{u}_n]} \sum_{u \in \mathcal{U}} (\widetilde{P}_{\text{Load}}^{u,t} - P_{\text{Load}}^{u,t})^{2}}}{\sqrt{\sum_{t\in [t^{u}_m,t^{u}_n]} \sum_{u \in \mathcal{U}} (P_{\text{Load}}^{u,t})^{2}}}\label{eq27},\\
  & \text{RE}_{\text{PV}} = \frac{\sqrt{\sum_{t\in [t^{u}_m,t^{u}_n]} \sum_{u \in \mathcal{U}} (\widetilde{P}_{\text{renew}}^{u,t} - p_{\text{renew}}^{u,t})^{2}}}{\sqrt{\sum_{t\in [t^{u}_m,t^{u}_n]} \sum_{u \in \mathcal{U}} (p_{\text{renew}}^{u,t})^{2}}}\label{eq28},
\end{align}
where $\widetilde{P}_{\text{Load}}^{u,t}$ and $\widetilde{P}_{\text{renew}}^{u,t}$ represent predicted load and PV generation for each prosumer $u$ in time slot $t$.

\section{Simulations and Discussions}\label{Simulation and Discussion}
\begin{figure}[!t]
\centering
\includegraphics[width=\columnwidth]{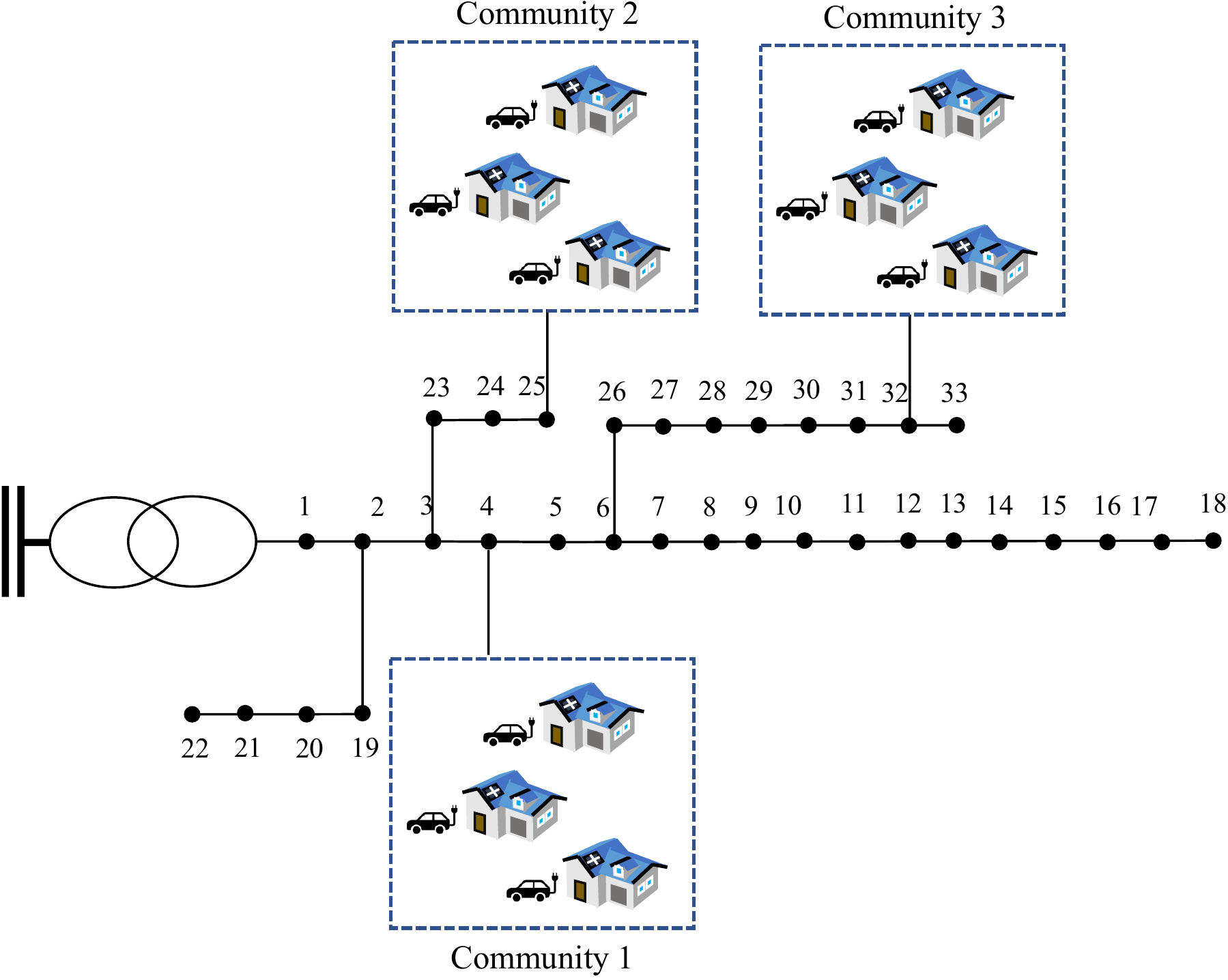}
\caption{IEEE 33 bus radial distribution test system with three communities of households and EVs.}
\label{fig:IEEE33bus}
\end{figure}

We evaluate our proposed network-aware value-stacking on the IEEE 33-bus test system, depicted in Fig. \ref{fig:IEEE33bus}, where three communities connect to nodes 4, 25, and 32, respectively. The total load of the IEEE 33-bus is $3.72$MW, and the base case values for the apparent power and voltage are $10$MVA and $12.66$kV, respectively. In each community, there are 20 prosumers (i.e. 20 households, each of which has a PV system and an EV). The EV battery capacity for each prosumer is $50$kWh, the initial stored energy ranges from $20$kWh to $30$kWh, and the maximum charging rate is $7$kW. We use real market data from NEM, ISO-NE, NY-ISO, and evaluate two tariffs for prosumers, e.g., TOU and TPT. For the TOU tariff, the off-peak price is AU\$ 0.2, and the peak price is AU\$ 0.32. For the TPT tariff, the energy price is AU\$ 0.2, and the peak price is AU\$ 0.8 \cite{tariff}. For the TOU plan in NY-ISO and ISO-NE, the off-peak price is US\$ 0.02, the shoulder price is US\$ 0.11, and the peak price is US\$ 0.32\footnote{Note that our evaluation of the EV value-stacking framework is based on cost reduction (\%) in each corresponding market, so different currency units (in AU\$ and US\$) do not affect our evaluation.}. The value-stacking model is validated in MATLAB using the YALMIP toolbox \cite{Yalmip} on a PC with Intel(R) Core(TM) i7-7700HQ CPU and 32GB RAM. 

\begin{figure}[!t]
\centering
\includegraphics[width=0.9\columnwidth]{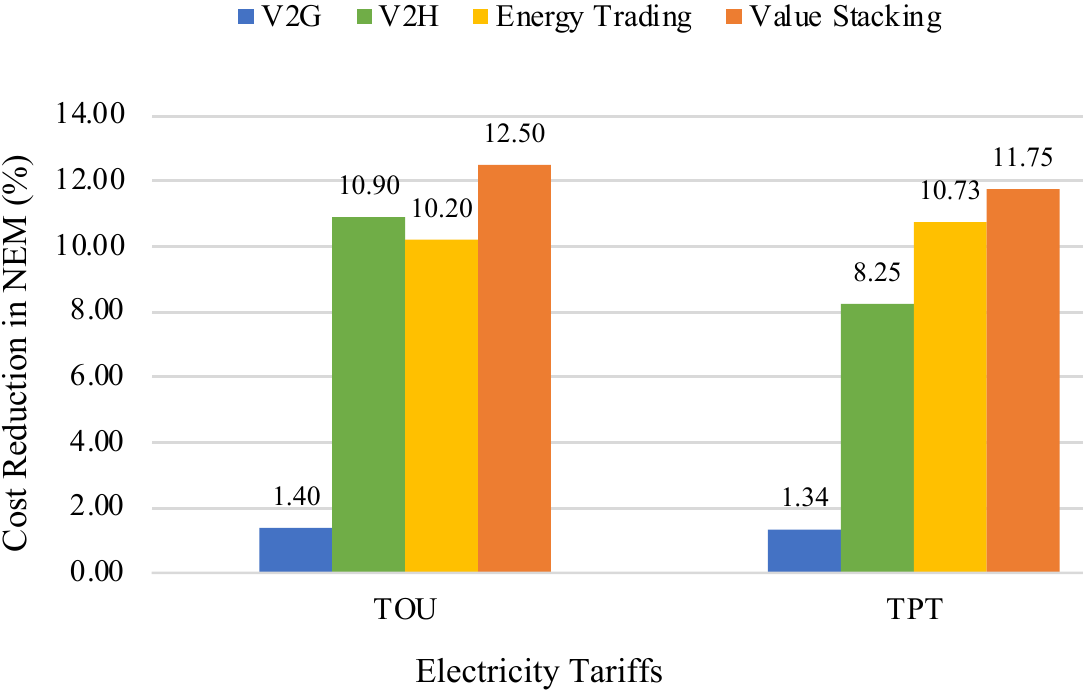}
\caption{Prosumers' cost reductions under two tariffs in four scenarios.}
\label{fig:cost reduction}
\end{figure}

\begin{figure}[!t]
\centering
\includegraphics[width=0.9\columnwidth]{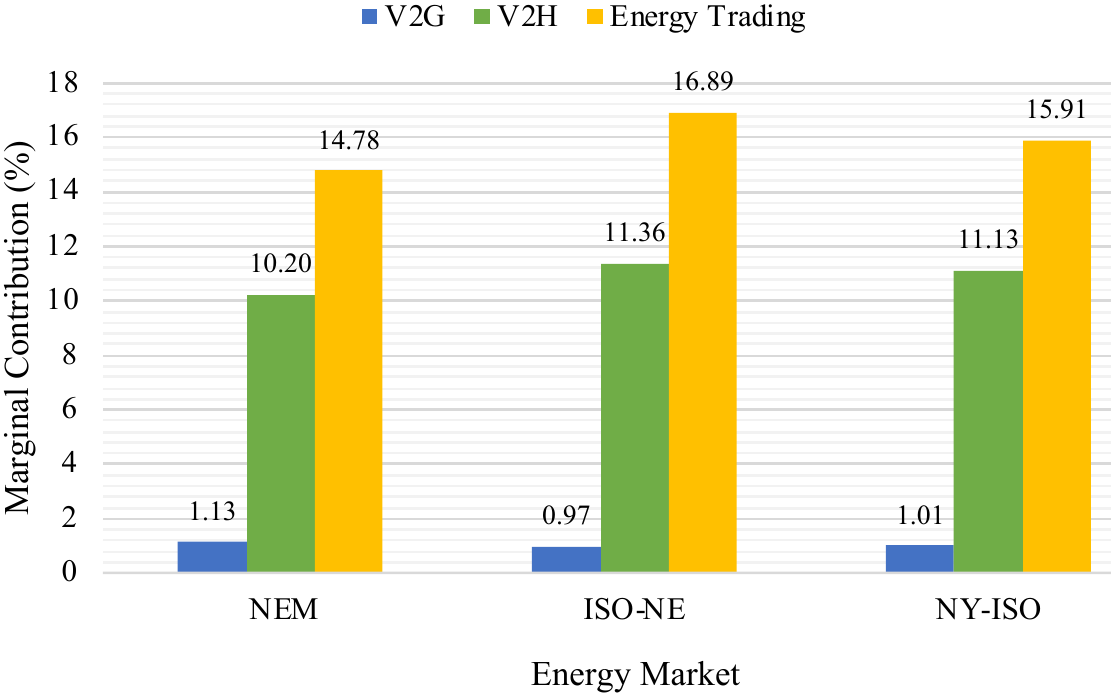}
\caption{Marginal contribution of different value streams in three markets.}
\label{fig:Marginal}
\end{figure}

\subsection{Value Stacking Performance Based on Perfect Prediction}

\begin{figure*}[!t]
\centering
\includegraphics[width=1.2\columnwidth]{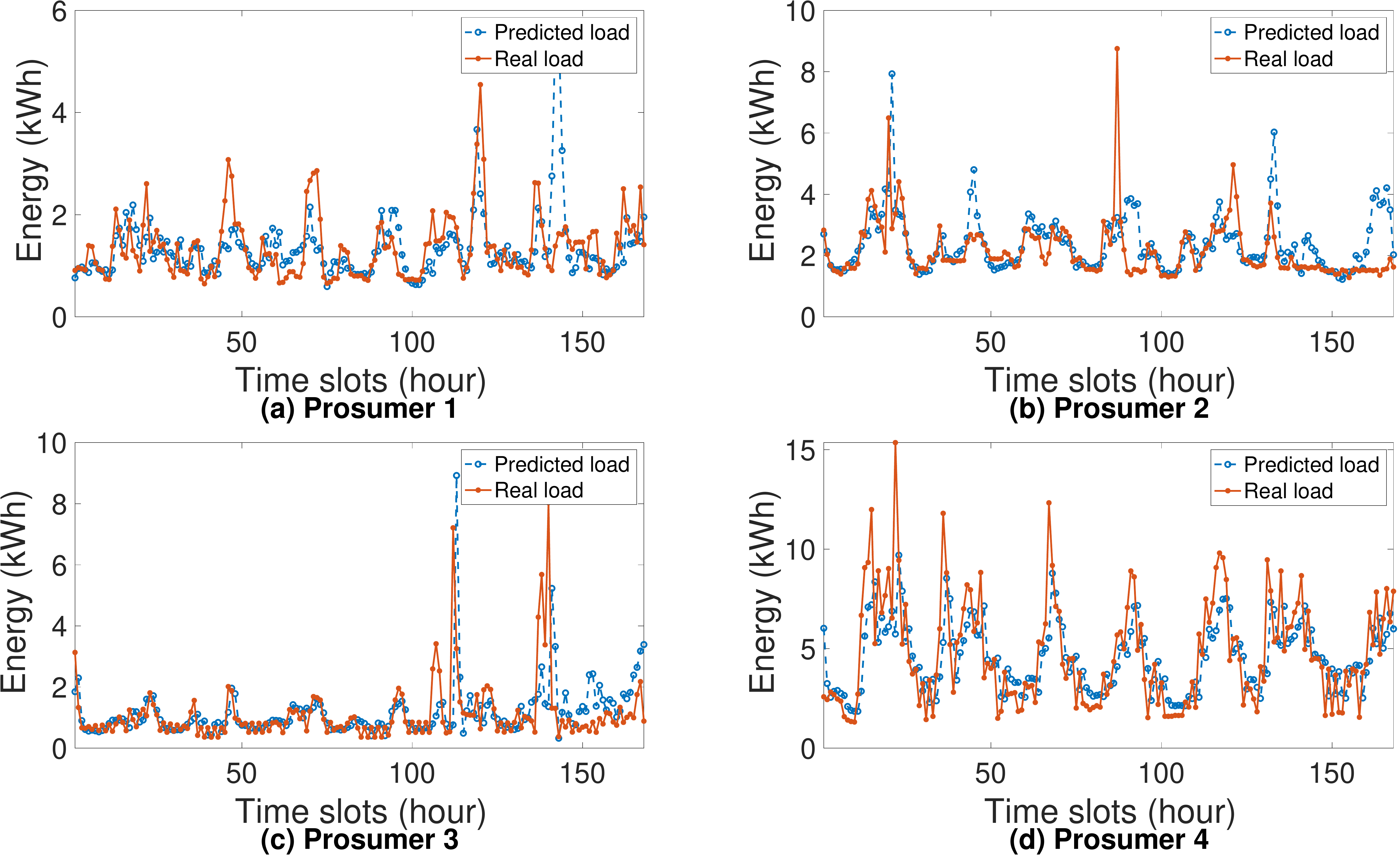}
\caption{Comparison of forecast load and real load for four prosumers.}
\label{fig:load forecast}
\end{figure*}

\begin{figure*}[!t]
\centering
\includegraphics[width=1.2\columnwidth]{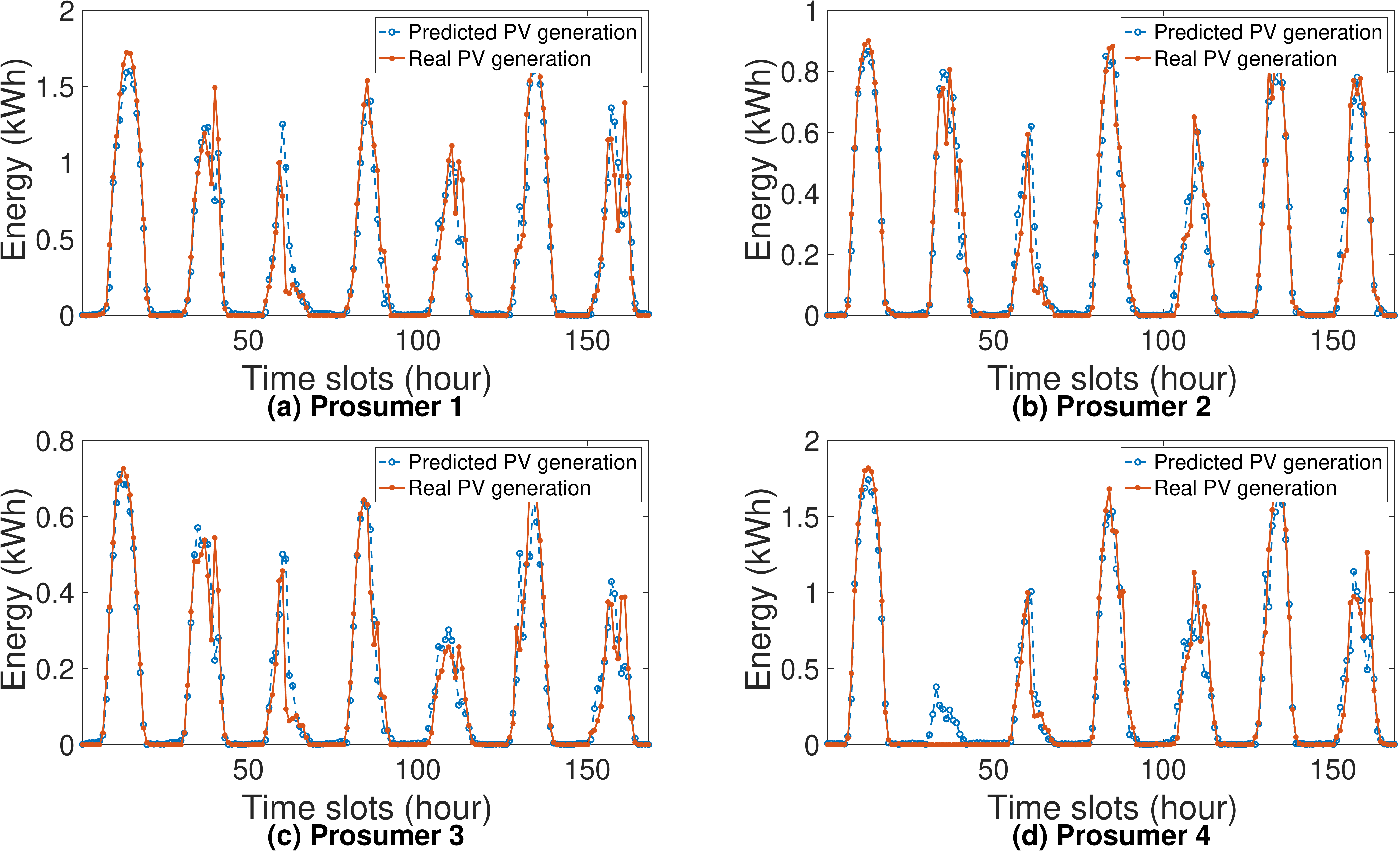}
\caption{Comparison of forecast PV generation and real PV generation for four prosumers.}
\label{fig:PV forecast}
\end{figure*}

We first evaluate the performance of value-stacking in terms of cost reduction for prosumers assuming perfect prediction of PV generation and residential load. The simulation horizon is 24 hours and we calculate the average daily cost over one week, including five working days and two holidays. The results are presented in Fig. \ref{fig:cost reduction}. The vertical axis represents the total cost reductions in $\%$ for value stacking and three baselines, including V2G, V2H, and EV energy trading alone, respectively. The two clusters of bars on the horizontal axis correspond to two tariffs, in which TOU and TPT are applied to prosumers, respectively. Under TOU pricing, value stacking achieves a cost reduction of $12.50\%$. Among three value streams, V2G creates the lowest cost reduction of $1.40\%$. The benefit of V2H is higher than energy trading in the local market since all prosumers have the same peak load hours, which leads to fewer opportunities for energy trading but more contribution of V2H in the cost reduction. Under TPT, value stacking achieves a cost reduction of $11.75\%$, which is lower compared to that under TOU pricing. Energy trading contributes more cost reduction than V2H because prosumers' peak loads can appear at diverse times and energy trading among prosumers can enable better peak shaving.

To evaluate the marginal contribution of each value stream in our value-stacking problem presented in Section~\ref{valuestacking}, we calculate the marginal contribution of each value stream in different markets, namely NEM, ISO-NE, and NY-ISO. The marginal contribution is computed as the relative difference of cost reductions between value stacking and baseline problems excluding one value stream. As Fig. \ref{fig:Marginal} shows, the marginal contribution of energy trading is much higher than those of the V2G and V2H in three markets, reaching up to $14.78\%$, $10.20\%$, and $1.13\%$, respectively. This indicates that prosumers can obtain the most cost reduction from energy trading when introduced in these three markets. Additionally, since the energy trading from the local market could better allocate energy for the value-stacking model to reduce the total cost of the whole system, prosumers prefer to sell power to the local market in the peak hour of electricity consumption rather than supply power for V2H, resulting in a minor contribution of V2H in three markets. In NY-ISO and NE-ISO, since the off-peak price of the TOU pricing is much lower than the peak and shoulder prices of the tariff, prosumers prefer to store energy in off-peak hours and sell energy to the local market (for energy trading) when the TOU tariff price is high in peak hours. Hence, energy trading benefits prosumers more in both NY-ISO and NE-ISO.

\subsection{Forecast Accuracy Assessment}
To forecast the household load and PV generation for prosumers, we utilize the LSTM network model in \ref{LSTM} to predict hourly load and PV generation for 60 prosumers, respectively, over $200$ days. Specifically, we adopt 12 months of data (from Jan. 1, 2020, to Dec. 31, 2020) of load and PV generation for prosumers as the training datasets, where the loss function is the root mean square error (RMSE). The number of hidden units is 400, the initial learning rate is 0.005, the maximum epochs is 250, and the Adaptive Moment Estimation (Adam) algorithm is utilized to train the LSTM model. Then, we forecast 200-day data (from 1/1/2021 to 19/7/2021) using the trained models. 

To evaluate the performance of the LSTM model, we randomly choose four prosumers during one week (i.e., 168 hours) to compare the load forecast with real load in Fig.~\ref{fig:load forecast}. According to Eq.~(\ref{eq27}), the relative errors in the predicted load for four prosumers are 0.47, 0.43, 0.74, and 0.36, respectively. For four prosumers, the prediction is not good enough in peak load hours, as it is challenging to capture the behavior of individual households. Prosumer 3 in Fig.~\ref{fig:load forecast} (c) has the largest relative error among the four prosumers since the error of forecast in peak load hour is large. Similar to load forecast, we also show the PV generation prediction for these four prosumers depicted in Fig.~\ref{fig:PV forecast}. Based on Eq.~(\ref{eq28}), we compute the relative errors of the predicted PV generation for four prosumers and the results are 0.25, 0.21, 0.26, and 0.22, respectively. Compared with load forecast, PV generation prediction achieves better performance, using the same LSTM architecture, which indicates that capturing the PV generation is relatively easier than household energy consumption behavior.

\subsection{Impact of Prediction Errors on Value Stacking Performance}
\begin{figure}[!t]
\centering
\includegraphics[width=0.8\columnwidth]{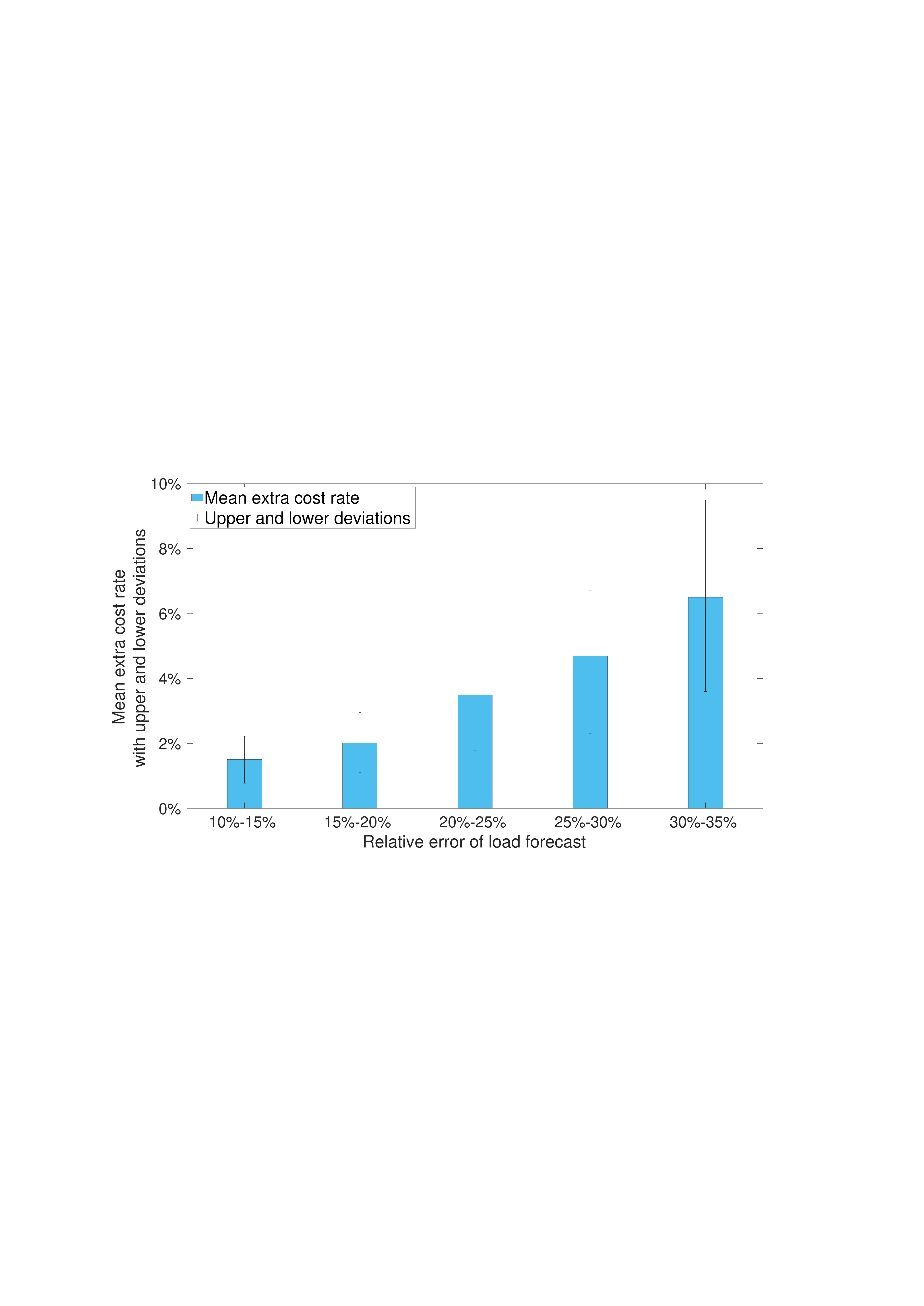}
\caption{Relationship between the mean extra cost rate and the relative error of load prediction.}
\label{fig:extra cost on load}
\end{figure}

\begin{figure}[!t]
\centering
\includegraphics[width=0.8\columnwidth]{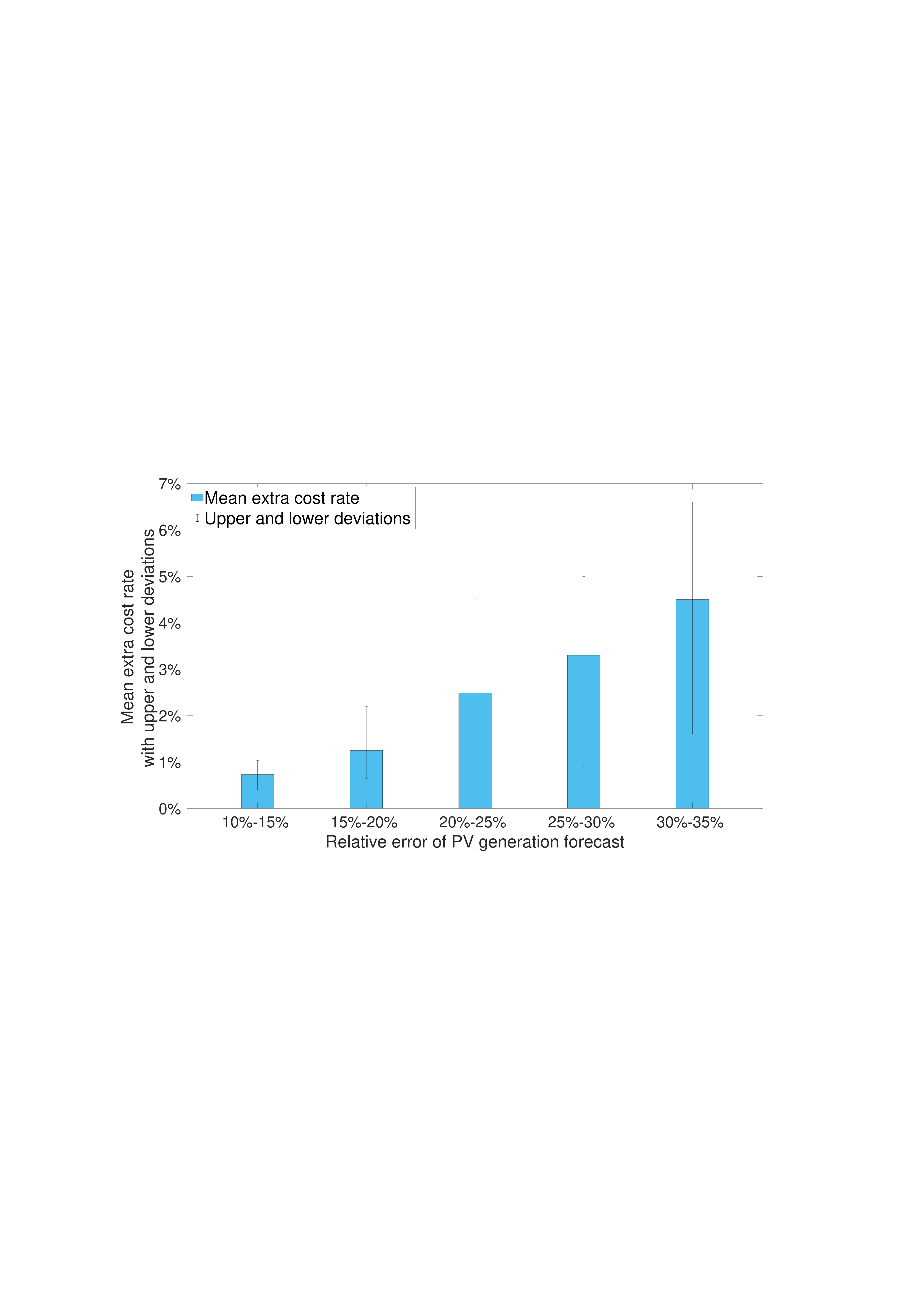}
\caption{Relationship between the mean extra cost rate and the relative error of PV generation prediction.}
\label{fig:extra cost on PV}
\end{figure}

To evaluate the impact of prediction errors, we deploy the predicted value of residential load and PV generation to calculate the extra cost according to Eq. (\ref{eq26}), respectively. Fig. \ref{fig:extra cost on load} shows the relationship between the relative error of load prediction and the mean extra cost rate (i.e., the expected averaged extra cost compared to the minimum cost) with upper and lower standard deviations. The mean extra cost rate increases as the load prediction error increases, achieving a mean of $6.5$\% ranging $[3.5,9.4\%]$ when the relative load prediction error is within $30 \sim 35$\%. 
Fig. \ref{fig:extra cost on PV} shows the relationship between the relative error of PV generation forecast and the mean extra cost rate. We see that the mean extra cost rate increases slightly as the relative error of the predicted PV generation increases. Compared with the impact of the load prediction errors, the impact of PV generation prediction errors is smaller. Specifically, when the relative PV prediction error achieves $30 \sim 35$\%, the mean extra cost rate is $4.4$\% ranging $[2.4,7.2\%]$. The results are obtained from our simulation setting and can be explained as follows. Sometimes, EVs are not at home in the daytime when the PV systems generate energy, and thus the impact of PV generation prediction errors on EV value-stacking is smaller.

\section{Conclusion}\label{Conclusion and Future Work}
In this paper, we proposed a V2X value-stacking optimization problem for EV coordination considering local network constraints to maximize the economic benefits. We formulated a RHO problem to solve the value-stacking problem in real-time using predicted PV generation and load information. We further evaluated the impact of prediction errors in load and PV generation on our proposed EV value-stacking problem. The results showed that the value-stacking model achieved better performance with a cost reduction of 12.50\%, when TOU was applied as the retail tariff. The marginal contributions of three value streams in three markets, namely NEM, ISO-NE, and NY-ISO, were also studied. The results demonstrated that energy trading contributed the most to cost reduction in three markets. The numerical results revealed that the mean extra cost was sensitive to the relative error in load prediction, while PV generation prediction error has a smaller impact. Our observation indicates that more accurate load prediction is needed for future EV value stacking models or systematic approaches are required to mitigate the impact of load prediction errors in value stacking optimization.

\bibliographystyle{IEEEtran}
\bibliography{ref.bib}

\end{document}